\documentclass[conference]{IEEEtran}
\usepackage{cite}
\usepackage{amsmath,amssymb,amsfonts}
\usepackage{algorithmic}
\usepackage{graphicx}
\usepackage{textcomp}
\usepackage{xcolor}
\def\BibTeX{{\rm B\kern-.05em{\sc i\kern-.025em b}\kern-.08em
    T\kern-.1667em\lower.7ex\hbox{E}\kern-.125emX}}

\usepackage{subcaption}
\usepackage{multicol}
\usepackage{booktabs}
\usepackage{subcaption}
\usepackage{url}
\usepackage{booktabs}
\usepackage{makecell}
\usepackage{arydshln}
\usepackage{multirow}
\usepackage{pgfplots}
\usepackage{hyperref}
\usepackage{float}
\usepackage{csquotes}
\usepackage{listings}
\usepackage{inconsolata}
\usepackage{tikz}
\usepackage{balance}

\pgfplotsset{compat=1.7}
\newcommand*\circled[1]{\tikz[baseline=(char.base)]{
            \node[shape=circle,draw,inner sep=0.5pt] (char) {#1};}}
            
\hypersetup{bookmarksnumbered=true,bookmarksopen=true}

\begin{document}
\title{SESR-Eval: Dataset for Evaluating LLMs in the
Title-Abstract Screening of Systematic Reviews
}

\IEEEoverridecommandlockouts
\IEEEpubid{\makebox[\columnwidth]{979-8-3315-9147-2/25/\$31.00 ©2025 IEEE \hfill}
\hspace{\columnsep}\makebox[\columnwidth]{ }}

\author{\IEEEauthorblockN{1\textsuperscript{st} Aleksi Huotala}
\IEEEauthorblockA{\textit{Department of Computer Science} \\
 \textit{University of Helsinki}\\
 Helsinki, Finland \\
 aleksi.huotala@helsinki.fi
 }
 \and
 \IEEEauthorblockN{2\textsuperscript{nd} Miikka Kuutila}
 \IEEEauthorblockA{\textit{Faculty of Computer Science} \\
\textit{Dalhousie University}\\
 Halifax, Canada \\
 miikka.kuutila@dal.ca
 }
 \and
 \IEEEauthorblockN{3\textsuperscript{rd} Mika Mäntylä}
 \IEEEauthorblockA{\textit{Department of Computer Science} \\
 \textit{University of Helsinki}\\
 Helsinki, Finland \\
 mika.mantyla@helsinki.fi
 }
 }
\maketitle
\IEEEpubidadjcol

\begin{abstract}
\textbf{Background:} The use of large language models (LLMs) in the title-abstract screening process of systematic reviews (SRs) has shown promising results, but suffers from limited performance evaluation.
\textbf{Aims:} Create a benchmark dataset to evaluate the performance of LLMs in the title-abstract screening process of SRs. Provide evidence whether using LLMs in title-abstract screening in software engineering is advisable.
\textbf{Method:} We start with 169 SR research artifacts and find 24 of those to be suitable for inclusion in the dataset. Using the dataset we benchmark title-abstract screening using 9 LLMs.
\textbf{Results:} We present the SESR-Eval (Software Engineering Systematic Review Evaluation) dataset containing 34,528 labeled primary studies, sourced from 24 secondary studies published in software engineering (SE) journals. Most LLMs performed similarly and the differences in screening accuracy between secondary studies are greater than differences between LLMs. The cost of using an LLM is relatively low -- less than \$40 per secondary study even for the most expensive model.
\textbf{Conclusions:} Our benchmark enables monitoring AI performance in the screening task of SRs in software engineering. At present, LLMs are not yet recommended for automating the title-abstract screening process, since accuracy varies widely across secondary studies, and no LLM managed a high recall with reasonable precision. In future, we plan to investigate factors that influence LLM screening performance between studies.
\end{abstract}


\section{Introduction}

Systematic reviews (SRs) are a research method used to identify and interpret relevant research on a particular topic ~\cite{kitchenhamGuidelinesPerformingSystematic2007}. This method is widely used in software engineering (SE) research. For example, more than a hundred SRs on software testing have been published to date~\cite{garousiSystematicLiteratureReview2016}. One critical step in the SR process is the screening of primary studies, where each candidate study is rigorously evaluated against a set of predefined criteria~\cite{kitchenhamGuidelinesPerformingSystematic2007}. The screening of primary studies in SRs typically involves two parts -- (1) title-abstract screening and (2) full-text screening. Title-abstract screening allows for rapid filtering of relevant primary studies. Studies that are clearly irrelevant can be excluded, and the included studies can then be screened in-depth by full-text screening ~\cite{kitchenhamGuidelinesPerformingSystematic2007}. This step is both time-consuming and prone to error, which has motivated recent research on the automation of the primary study screening process using large language models (LLMs) in SE ~\cite{huotalaPromiseChallengesUsing2024a, felizardoChatGPTApplicationSystematic2024} and more extensively in biomedical and medical domains~\cite{liEvaluatingEffectivenessLarge2024,  dennstadtTitleAbstractScreening2024,tranSensitivitySpecificityAvoidable2023, delgado-chavesTransformingLiteratureScreening2025, matsuiHumanComparableSensitivityLarge2024a, guoAutomatedPaperScreening2024b}.

While prior studies show promising results for title-abstract screening in SE ~\cite{huotalaPromiseChallengesUsing2024a, felizardoChatGPTApplicationSystematic2024}, they suffer from a small number of secondary studies and LLMs under evaluation. One study~\cite{huotalaPromiseChallengesUsing2024a} evaluated performance of two LLMs in a single secondary study, while another study~\cite{ felizardoChatGPTApplicationSystematic2024} evaluated performance of a single LLM with two secondary studies. Performance evaluations based on small datasets risk introducing bias, making it difficult to assess the utility of the solution.

Clearly, there is a need for a larger benchmarking dataset for the title-abstract screening task in SE. Only in this way can we monitor LLM performance and provide evidence-based advice on whether or not to use LLMs when conducting systematic reviews. 
To address this gap, we constructed a larger benchmark dataset using the research artifacts of 24 SE secondary studies and evaluated the title-abstract screening performance of 9 different LLMs.

\def\firstrq {Can we create a benchmark dataset from research artifacts to evaluate the performance of title-abstract screening?}
\def\secondrq {What is the performance of the LLMs in the screening task?}
\def\thirdrq {How does screening performance vary across secondary studies?}
\def\fourthrq {What is the combined effect of LLMs and secondary studies on the screening results?}
\def\fifthrq {How does the time, token efficiency, and cost vary across different LLMs?}

\section{Background}
\label{section:background}

\subsection{Automating the Screening Process of Systematic Reviews}

\begin{table}
    \centering
    \caption{Related work that evaluate title-abstract screening in different domains and evaluation methods with LLMs.}
    \label{table:priorworks}
\resizebox{\linewidth}{!}{%
    \begin{tabular}{crrcrr}
        \toprule
         \textbf{Study} & \textbf{Domain} & \textbf{LLMs} & \textbf{\makecell[c]{Evaluation\\method}} & \textbf{\makecell[r]{Secondary\\studies}} & \textbf{\makecell[r]{Primary\\studies}} \\
         \midrule
         
         \cite{huotalaPromiseChallengesUsing2024a} & SE & 2 & Binary & 1 & 1,306 \\
         \cite{felizardoChatGPTApplicationSystematic2024} & SE & 1 & Likert & 2 & 582 \\
         \hdashline
         \cite{liEvaluatingEffectivenessLarge2024} & Biomed & 9 & Binary & 3 & 505 \\
         \cite{dennstadtTitleAbstractScreening2024} & Biomed & 3 & Likert & 10 & 38,426 \\
         \hdashline
         \cite{tranSensitivitySpecificityAvoidable2023} & Med & 1 & Binary & 5 & 22,666 \\
         \cite{delgado-chavesTransformingLiteratureScreening2025} & Med & 18 & Binary & 3 & 6,217\\
         \cite{matsuiHumanComparableSensitivityLarge2024a} & Med & 2 & Binary & 2 & 4,527 \\
         \cite{guoAutomatedPaperScreening2024b} & Med & 2 & Binary & 6 & 24,307 \\
        \midrule
         \textbf{Ours} & \textbf{SE} & \textbf{9} & \textbf{Both} & \textbf{24} & \textbf{34,528} \\
         \bottomrule
    \end{tabular}
}%
\end{table}

Table \ref{table:priorworks} summarizes related work in this area. In the SE domain, two studies~\cite{huotalaPromiseChallengesUsing2024a, felizardoChatGPTApplicationSystematic2024} investigated the use of LLMs for screening with two different evaluation methods and 1,888 primary studies in total. In contrast, the biomedical and medical domains have seen more research on the topic~\cite{liEvaluatingEffectivenessLarge2024, dennstadtTitleAbstractScreening2024, tranSensitivitySpecificityAvoidable2023, delgado-chavesTransformingLiteratureScreening2025,matsuiHumanComparableSensitivityLarge2024a,guoAutomatedPaperScreening2024b}. Studies outside of SE have used significantly larger sample of both secondary and primary studies.

\subsection{Benchmarking Large Language Models}

\begin{table}[tbp]
    \centering
    \caption{Commonly used datasets to benchmark LLMs.}
    \label{table:llm_datasets}
\begin{tabular}{lr}
\hline
\toprule
\textbf{Name} & \textbf{Domain} \\
\midrule
AGIEval \cite{liuAreLLMsGood2024, zhongAGIEvalHumanCentricBenchmark2023} & Knowledge \& reasoning \\
ARC-Challenge \cite{clarkThinkYouHave2018} & Knowledge \& reasoning \\
BIG-Bench-Hard \cite{suzgunChallengingBIGBenchTasks2023} & Knowledge \& reasoning \\
GPQA \cite{reinGPQAGraduateLevelGoogleProof2023}, Diamond & Knowledge \& reasoning \\
HellaSwag \cite{zellersHellaSwagCanMachine2019} & Knowledge \& reasoning \\
MMLU \cite{liuAreLLMsGood2024, hendrycksMeasuringMassiveMultitask2021} & Knowledge \& reasoning \\
\hdashline
GSM8K \cite{liuAreLLMsGood2024, cobbeTrainingVerifiersSolve2021} & Mathematical \& logical reasoning \\
MATH \cite{liuAreLLMsGood2024, hendrycksMeasuringMathematicalProblem2021} & Mathematical \& logical reasoning \\
MathVista (testmini) \cite{luMathVistaEvaluatingMathematical2024} &  Mathematical \& logical reasoning \\
MGSM \cite{shiLanguageModelsAre2022} & Mathematical \& logical reasoning \\
MMMU \cite{yueMMMUMassiveMultidiscipline2023} & Mathematical \& logical reasoning \\
\hdashline
HumanEval \cite{liuAreLLMsGood2024, chenEvaluatingLargeLanguage2021} & Programming \& code generation \\
MBPP \cite{austinProgramSynthesisLarge2021} & Programming \& code generation \\
WikiSQL \cite{zhongSeq2SQLGeneratingStructured2017} & Programming \& code generation \\
\hdashline
DROP \cite{duaDROPReadingComprehension2019} & Multimodal \& visual reasoning \\
AI2D, test \cite{kembhaviDiagramWorthDozen2016} & Multimodal \& visual reasoning  \\
ChartQA \cite{masry-etal-2022-chartqa} & Multimodal \& visual reasoning \\
MMBench \cite{liuMMBenchYourMultimodal2024} & Multimodal \& visual reasoning \\
\midrule
\textbf{Ours:} SESR-Eval & Primary study title-abstract screening \\
\bottomrule
\end{tabular}
\end{table}

LLMs are used in various domains. To verify the LLMs' performance and accuracy, benchmarks have been created for these domains. Table \ref{table:llm_datasets} lists some of the most popular datasets for benchmarking LLMs. The list covers domains ranging from knowledge and reasoning, mathematical reasoning, code generation to multimodal and visual reasoning. More datasets for additional domains can be found online, from the "Papers with Code"-website\footnote{\url{https://paperswithcode.com/datasets}}. As LLMs are used more broadly, benchmark datasets must become larger and more diverse to assess their performance reliably. Reasoning tasks range from common knowledge to context-specific tasks, which motivates the creation of datasets for specific types of tasks, such as title-abstract screening in SRs. To the best of our knowledge, no public datasets currently exist for benchmarking title-abstract screening.

\begin{figure*}[tbp]
\resizebox{\textwidth}{!}{%
\includegraphics{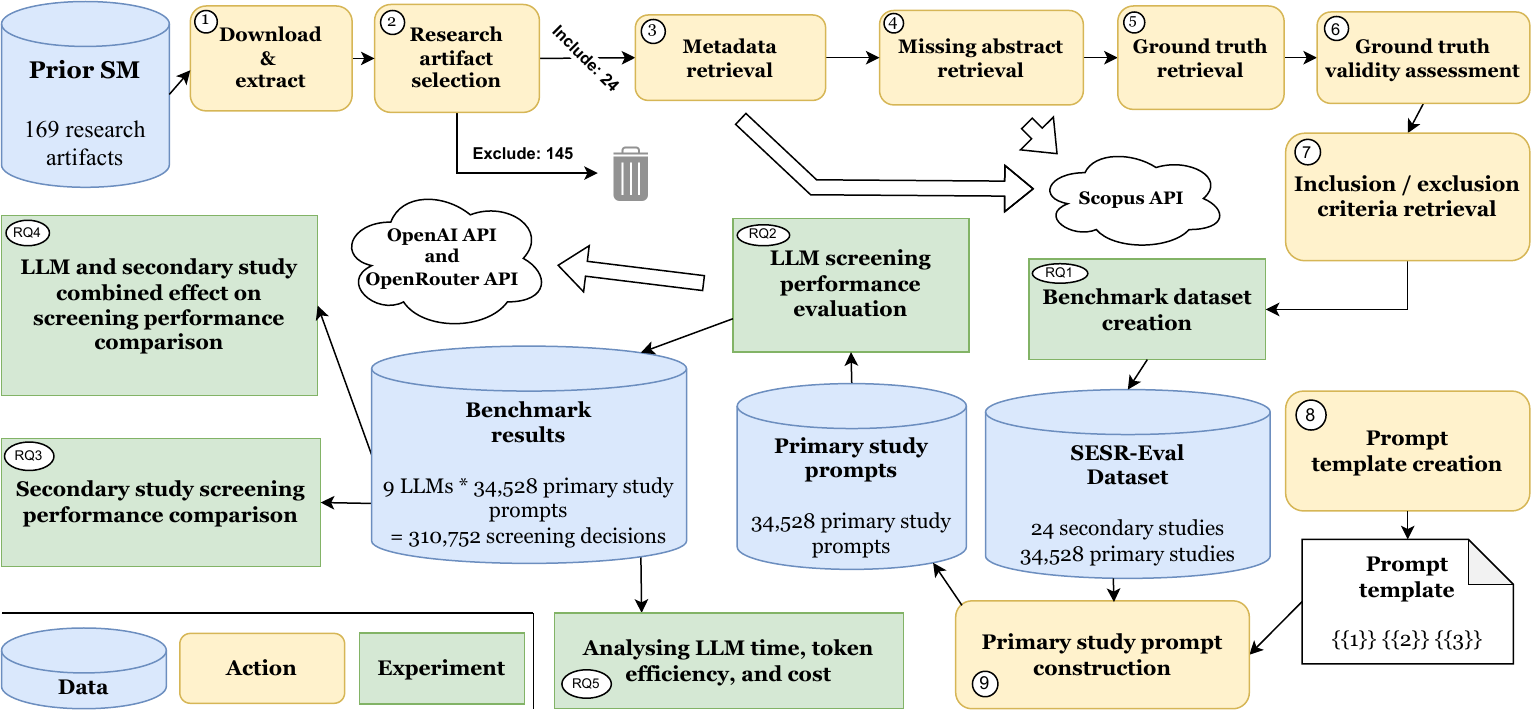}
}%
\caption{The dataset creation process, including numbered actions and the conducted experiments.}
\label{figure:dataset_creation}
\end{figure*}

\section{Methods}
\label{section:methods}

Fig. \ref{figure:dataset_creation} shows an overview of how the dataset was created and how the experiments used to answer our research questions were conducted. We followed the SIGSOFT Empirical Standards for Software Engineering Research~\cite{ralph2020empirical} as a framework for conducting this study. These standards provide guidance on reporting benchmarking studies.

\subsection{Research questions}
The following research questions were formulated to study the creation of the benchmark dataset and evaluate the screening performance of LLMs in the title-abstract screening task:

\begin{itemize}
    \item \textbf{RQ1:} \textit{\firstrq} 
    Motivation: Primary study screening is a laborious task. A benchmark dataset can be used to provide evidence on whether LLMs are useful for this task. Furthermore, a dedicated dataset enables comparing AI-based solutions and allows us to monitor the progress and improvement of LLMs in the future.
    \item \textbf{RQ2:} \textit{\secondrq}
    Motivation: This research question provides evidence on where state-of-the-art LLMs currently stand.
    \item \textbf{RQ3:} \textit{\thirdrq}
    Motivation: It is likely that the performance of the LLM is not only a function of the model's quality, but also a function of the SR itself. Therefore, it is important to compare how screening performance varies across secondary studies. Secondly, evaluating LLM performance across secondary studies highlight how well LLMs understand inclusion and exclusion criteria and the domain of the study.
    \item \textbf{RQ4:} \textit{\fourthrq}
    Motivation: Understanding the combined effect of LLMs and secondary studies is important if certain LLMs perform significantly better in screening on specific secondary studies.
    \item \textbf{RQ5:} \textit{\fifthrq}
    Motivation: The high token usage, latency, or cost of LLMs may discourage researchers from using them. The paper title-abstract screening process can involve thousands of studies, which quickly raises token usage, cost, and processing time. By studying how time, token efficiency, and cost varies across different LLMs, we can make recommendations to researchers about which models offer the best trade-off between performance and latency.
\end{itemize}

\subsection{Creating the benchmark dataset}
\label{section:methods:datasetcreation}
\label{subsection:dataset}

This subsection provides an answer for \textbf{RQ1}. Fig. \ref{figure:dataset_creation} visualizes the benchmark dataset creation process. We used a prior systematic mapping~\cite{huotalaReplicationPackagesSoftware2025} as our source for research artifacts, which lists 169 research artifacts.

\begin{table}[tbp]
    \centering
    \caption{Data extraction format used in the research artifact selection.}
    \label{table:analysis_format_extraction}
    \begin{tabular}{lr}
    \toprule
        \textbf{Item} & \textbf{Type}\\
        \midrule
        Initial papers & Boolean \\
        Included papers & Boolean \\
        Excluded papers &  Boolean \\
        Selection criteria & Boolean \\
        Data extraction & Boolean \\
        Screening results & Boolean \\
        Notes & Text \\
        Dataset size & Numeric \\
        Papers w/o title & Numeric \\
        Papers w/o abstract & Numeric \\
        Papers w/o keywords & Numeric \\
        License & Text\\
        \bottomrule
    \end{tabular}
    \label{table:data_extraction_and_analysis_format}
\end{table}

The first step of creating the benchmark dataset was to download and extract the research artifacts (\circled{1} in Figure \ref{figure:dataset_creation}). Next, for the extracted artifacts, the research artifact selection was conducted (\circled{2}). The research artifact selection consisted of data extraction in a format specified in Table~\ref{table:data_extraction_and_analysis_format} and the following checks:

\begin{enumerate}
    \item The secondary study performs title-abstract screening.
    \item The research artifact contains title-abstract screening data.
    \item The title-abstract screening data is in a suitable format.
\end{enumerate}

From the data extraction (Table \ref{table:data_extraction_and_analysis_format}), the "Screening results" information is used to include or exclude research artifacts. We double-coded 20 randomized research artifacts regarding whether the secondary study contains screening data and that data is suitable for our experiment. We had an agreement of 90\% and 
 95\%, corresponding to Krippendorff's alphas~\cite{krippendorff2011computing} of $0.786$ and 
 $0.831$ respectively. The rest of the artifacts were then investigated by the first author.
 
The data in the research artifacts may span across multiple files or be in one file. If it was not obvious that the research artifact contains title-abstract screening data, we studied the secondary study research paper to find out if there is information regarding the screening data.
 Through this process we found multiple reasons why we excluded research artifacts, including: 
\begin{itemize}
    \item The secondary study did not perform title-abstract screening.
    \item The secondary study had significant ambiguities in title-abstract screening. For instance, the research artifact didn't differentiate title-abstract screening from full-text screening.  
    \item The research artifact was partial or had omissions, which made primary study retrieval impossible -- e.g., only partial titles without other identifiers were given in the research artifact.
    \item The research artifact was inaccessible -- e.g., links to dead webpages or pages requiring authentication.
\end{itemize}

After the research artifact selection was completed, we conducted metadata retrieval (\circled{3}). Metadata retrieval was done with the Scopus API\footnote{\label{footnote:scopus}. The data was downloaded from Scopus API between January 1 and 18 July, 2025 via http://api.elsevier.com and http://www.scopus.com.} to retrieve relevant metadata -- e.g., primary study DOI, keywords and authors. We deem this metadata retrieval a necessary action for providing a high-quality benchmark dataset, where additional metadata can be later retrieved by the paper's DOI or in our instance -- Scopus EID. 

After the primary study metadata was retrieved, we retrieved abstracts (\circled{4}) for the primary studies missing them. Missing abstracts were retrieved by the following steps:

\begin{enumerate}
    \item Automatically from Scopus \footref{footnote:scopus}
    \item Manually with Google Scholar, if Step (1) failed to retrieve the abstract and if the secondary study has fewer than 200 missing primary study abstracts \footnote{Manual abstract retrieval is slow, thus, we had to limit the amount of abstracts we manually retrieved.} 
\end{enumerate}

After the missing abstract retrieval, we retrieved the ground truth (\circled{5}) by looking at the research artifact data and determining the correct label for the data. Ground truth is the screening decision of the primary study, made by human researcher experts. Retrieving the ground truth involved full-text reading of the secondary study to understand, how the screening decision was described in the study and the research artifact. 

After the ground truth was retrieved, we manually assessed their correctness (\circled{6}) - by once again full-text reading the secondary study and checking if the screening phase paper counts are equal in the secondary study and in the research artifact. This guarantees a rigorous evaluation of the criteria.

After the ground truth validity assessment, we retrieved the inclusion and exclusion criteria from the secondary study (\circled{7}). This information is necessary for title-abstract screening, as these are the fundamental rules, which the actual ground truth is based upon.

After the criteria retrieval, we generated unique identifiers for each secondary study in the dataset to distinguish them when running benchmarks. Finally, we combined the primary studies from each secondary study to form the final SESR-Eval dataset. Papers with missing titles or abstracts were excluded from the final dataset, as title-abstract screening is meant to be conducted with non-empty titles and abstracts. The final dataset is available in the research artifact of this study~\cite{zenodo_2_anon}.

\subsection{Evaluating the screening performance of the LLMs}
\label{section:evaluation_llm_performance}

This subsection provides background for research questions \textbf{RQ1}, \textbf{RQ2} and \textbf{RQ3}.

\subsubsection{Creating the zero-shot prompt template and constructing the primary study prompts}

\begin{figure}[tbp]
\vspace*{-0.3\baselineskip}
\begin{minipage}{\linewidth}
\begin{lstlisting}[breaklines=true, aboveskip=0pt, belowskip=0pt]
Role: You are a software engineering researcher conducting a systematic literature review (SLR).

Task: Evaluate a primary study using **three types of assessments**, applied to both:
a) The **overall** relevance of the primary study
b) Each individual **inclusion/exclusion criterion**

### Assessment Types:
1) **Binary classification**
    - **Value:** `"true"` or `"false"`
    - **Interpretation:** Whether the criterion or relevance is clearly met (true) or not (false).
2) **Probability classification**
    - **Value:** A float between `0.000` and `1.000`
    - **Interpretation:** The likelihood, that the criterion applies or the primary study is relevant.
        - A value closer to `1.000` means that it is extremely likely (very strong match)
        - A value closer to `0.000` means it is extremely unlikely (very weak or no match)
        - You are encouraged to use intermediate values (e.g. `0.100`, `0.250`, `0.350`, `0.700`, `0.950`, `0.999` etc..), not just `0.000` or `1.000`
3) **Likert scale**
    - **Value:** An integer from `1` to `7`
    - **Interpretation:** Degree of agreement with the criterion being met, or the relevance of the study
        - 1: Strongly disagree
        - 2: Disagree
        - 3: Somewhat disagree
        - 4: Neither agree nor disagree
        - 5: Somewhat agree
        - 6: Agree
        - 7: Strongly agree
### Important:
You **must provide all three types of assessments** for:
a) The overall relevance of the primary study
b) Each individual inclusion or exclusion criterion

### Inclusion and exclusion criteria:

### Additional instructions:

### Primary study:
**Title:**
**Abstract:**
\end{lstlisting}
\end{minipage}
\caption{Zero-shot prompt template.}
\label{fig:prompt_template}
\end{figure}

Before constructing the primary study prompts, we created the zero-shot prompt template (\circled{8}), in Fig. ~\ref{fig:prompt_template}. As a starting point for the prompt template, we used the zero-shot prompt from a SE paper, which studied title-abstract screening using LLMs~\cite{huotalaPromiseChallengesUsing2024a}. Next, we added Likert scale decisions, used by Felizardo et al.~\cite{felizardoChatGPTApplicationSystematic2024}, for each criterion (include or exclude) being met, and the relevance of the whole study: \begin{displayquote}1) Strongly disagree; 2) Disagree; 3) Somewhat disagree; 4) Neither agree nor disagree; 5) Somewhat agree; 6) Agree and 7) Strongly agree.\end{displayquote} Finally, we added a floating point decision, which has a value between $0.000$ and $1.000$. A score of $1.000$ means the paper (or criterion) is included with high probability; secondly - a score of $0.000$ means the paper (or criterion) is most likely excluded. Our goal was to give the LLM as detailed prompt as possible, to reduce the chance of hallucinations and improve the overall task performance. We detailed each assessment type and emphasized that the LLM is required to produce all three types of assessments for the overall relevance of the study and for each individual inclusion-exclusion criteria.

After the prompt template had been created, the primary study prompts were constructed (\circled{9}) by using every title-abstract pair and their corresponding inclusion-exclusion criteria from the SESR-Eval dataset. The result is a set of primary study prompts.

\begin{table}[tbp]
\centering
    \caption{LLMs used in the comparison.}
    \label{table:list_of_llms}
\resizebox{\linewidth}{!}{
    \begin{tabular}{lllcc}
    \toprule
    \textbf{Company} & \textbf{LLM \& Model} & \textbf{Parameters} & \textbf{Reasoning} & \textbf{\makecell[l]{Open\\source}} \\
    \midrule
    \multirow{2}{*}{OpenAI} & o3-mini$^{\circ}$ & \multirow{2}{*}{Unknown$^{**}$} & \multirow{2}{*}{Yes} & \multirow{2}{*}{No} \\
        & \textit{\texttt{o3-mini-2025-01-31}} &  &  &  \\
    \hdashline
    \multirow{2}{*}{OpenAI} & GPT-4o & \multirow{2}{*}{Unknown$^{**}$} & \multirow{2}{*}{No} & \multirow{2}{*}{No} \\
           & \textit{\texttt{gpt-4o-2024-11-20}} &  &  &  \\
    \hdashline
    \multirow{2}{*}{OpenAI} & GPT-4.1 & \multirow{2}{*}{Unknown$^{**}$} & \multirow{2}{*}{No} & \multirow{2}{*}{No} \\
     & \textit{\texttt{gpt-4.1-2025-04-14}} &  &  &  \\
    \hdashline
    \multirow{2}{*}{OpenAI} & GPT-4.1 mini & \multirow{2}{*}{Unknown$^{**}$} & \multirow{2}{*}{No} & \multirow{2}{*}{No} \\
     & \textit{\texttt{gpt-4.1-mini-2025-04-14}} &  &  &  \\
    \hdashline
    \multirow{2}{*}{OpenAI} & GPT-4.1 nano & \multirow{2}{*}{Unknown$^{**}$} & \multirow{2}{*}{No} & \multirow{2}{*}{No} \\
     & \textit{\texttt{gpt-4.1-nano-2025-04-14}} &  &  &  \\
    \hdashline
    \multirow{2}{*}{Anthropic} & Claude 3.7 Sonnet & \multirow{2}{*}{Unknown$^{**}$} & \multirow{2}{*}{No} & \multirow{2}{*}{No} \\ 
     & \textit{\texttt{anthropic/claude-3.7-sonnet}} &  &  &  \\
    \hdashline
    \multirow{2}{*}{DeepSeek} & DeepSeek R1 & \multirow{2}{*}{671B} & \multirow{2}{*}{Yes}  &   \multirow{2}{*}{Yes} \\
     & \textit{\texttt{deepseek/deepseek-r1}} &  &   &    \\
    \hdashline
    \multirow{2}{*}{Meta} & Llama 4 Maverick & \multirow{2}{*}{400B} & \multirow{2}{*}{No}  & \multirow{2}{*}{Yes} \\
     & \textit{\texttt{meta-llama/llama-4-maverick}} &  &   &  \\
    \hdashline
    \multirow{2}{*}{Mistral} & Ministral 8B & \multirow{2}{*}{8B} & \multirow{2}{*}{No} & \multirow{2}{*}{Yes}  \\
     & \textit{\texttt{mistral/ministral-8b}} &  &  &   \\
     \midrule
    \textbf{Configuration} & \multicolumn{4}{l}{temperature = $0.0$, top\_p = $0.1$} \\
    \bottomrule
    \\
    \multicolumn{5}{l}{\small $^{\circ}$ Reasoning effort is set to "high".} \\
    \multicolumn{5}{l}{\small $^{**}$ OpenAI and Anthropic do not disclose the parameter counts of their models.} \\
    \end{tabular}
}%
\end{table}

\subsubsection{Benchmarking title-abstract screening performance with LLMs}

To study the screening performance of the LLMs with the larger dataset (\textbf{RQ2}) and across secondary studies (\textbf{RQ3}), we benchmarked the title-abstract screening performance with LLMs listed in Table \ref{table:list_of_llms}. We aimed to include a diverse set of (a) reasoning, (b) non-reasoning, (c) commercial and (d) open-source LLMs, which publish their model weights or model source code online. For the reasoning models, we selected a commercial and an open-source LLM. Second, for non-reasoning models, we selected four commercial and two open-source LLMs. LLMs from these vendors are commonly used in SE research and are used in prior work~\cite{huotalaPromiseChallengesUsing2024a, felizardoChatGPTApplicationSystematic2024, houLargeLanguageModels2024, calciolariIntegratingLargeLanguage2025, delgado-chavesTransformingLiteratureScreening2025}. Similarly, as the performance of neural networks with a relatively small number of parameters, sometimes also called small language models (SMLs) can achieve competitive performance~\cite{schick2021s}, we decided to add two models that could be considered such to better understand time and token efficiency related to model size. In all of our benchmark experiments, we used the following system prompt: \begin{displayquote}"You are an expert research assistant.''\end{displayquote} With the LLMs, we used structured JavaScript Object Notation (JSON) response format instead of the traditional textual response. Using a structured response format allows for a predictable output token count, programmatic integration and faster evaluation of the LLM output~\cite{gengGeneratingStructuredOutputs2025,liuAreLLMsGood2024}. The response format we used is described in Table ~\ref{table:json_schema} and available in the research artifact~\cite{zenodo_2_anon}. 

\begin{table}[tbp]
\centering
\caption{JSON response format for the LLMs.}
\resizebox{\linewidth}{!}{
\begin{tabular}{ll}
\toprule
\textbf{Field} & \textbf{Description} \\
\midrule
overall\_decision & Final binary, probability and likert decision and reasoning \\
inclusion\_criteria & List of inclusion criteria and its binary, probability and likert decision \\
exclusion\_criteria & List of exclusion criteria and its binary, probability and likert decision \\
\bottomrule
\end{tabular}
}
\label{table:json_schema}
\end{table}

OpenAI LLMs were called directly using OpenAI's API endpoints. For the remaining LLMs, we used OpenRouter~\footnote{\url{https://openrouter.ai/}} - a service that provides an unified interface to test LLMs in a provider-agnostic way.

To compare the performance of title-abstract screening between LLMs and secondary studies (\textbf{RQ4}), we fit a logistic regression model predicting the correctness of screening decisions based on the LLM and the primary study. The logistic regression model was fitted and the odds ratios were calculated using the R programming language and the methods $glm$ and $coef$ included in the base R distribution version $4.3.3$. For calculating Krippendorff's Alpha, we used the R-package $irr$~\cite{gamer2012package} version 0.84.1. We also calculated the precision, recall and the F1 score for the tested LLMs and all primary studies and secondary studies from the benchmark set, using Python's scikit-learn library~\cite{pedregosa2011scikit}. The code for these is given in our research artifact.

For all LLMs, we set the temperature parameter~\footnote{\label{openai_footnote} \url{https://platform.openai.com/docs/api-reference/completions/create}} to zero and the top\_p\footref{openai_footnote} parameter to $0.1$ -- configuration used in related works~\cite{huotalaPromiseChallengesUsing2024a,felizardoChatGPTApplicationSystematic2024}. The temperature parameter controls how deterministic the LLM's output is. The lower the temperature value, the more the LLM attempts to always output the same tokens for the same prompt, regardless of the number of calls to the LLM. Secondly, the top\_p parameter controls the top n-\% tokens that are considered in the output. Setting top\_p to $0.1$ means that the LLM attempts to sample from the top 10\% most probable tokens, which guides the LLM to output more relevant tokens.

\subsection{\fifthrq}

This subsection provides background for the last research question \textbf{(RQ5)}. To evaluate the time, token efficiency and cost of the LLMs, we analyzed the latency and input and output token counts for each of the LLMs using a custom Python script. The latency and token information is returned by the LLMs' API. Latency was measured as the duration of sending a single request to the LLM and receiving its response. We used the 95th percentile~\cite{hyndmanSampleQuantilesStatistical1996} of response times to evaluate performance under typical conditions. The total token count of the LLM is a combination of the input (=prompt) and output tokens, where reasoning models are expected to output additional "reasoning tokens", which are aggregated to the output tokens. Providers, such as OpenAI, cache input tokens to reduce costs~\footnote{\url{https://platform.openai.com/docs/guides/prompt-caching}}. Costs were computed from the token count, multiplied by the LLMs price per 1M tokens. 

\section{Results}

\begin{table*}
	\centering
	\caption{List of secondary studies included in the dataset, with statistics.}
	\label{table:list_of_studies_with_screening_data}
	\resizebox{\linewidth}{!}{%
		\begin{tabular}{c|r:rlr:r|rccr}
			\toprule
			\textbf{Study}                                           & \textbf{\makecell[r]{Total \\ studies}} & \textbf{\makecell[r]{Included \\ studies}} & \textbf{\makecell[l]{Excluded \\ studies}} & \textbf{\makecell[r]{I/E \\ Ratio}} & \textbf{\makecell[r]{Missing \\ abstracts}} & \textbf{Journal} & \textbf{\makecell[c]{Inclusion \\ criteria}} & \textbf{\makecell[c]{Exclusion \\ criteria}} & \textbf{\makecell[r]{SWEBOK \\ Knowledge Area / Supplement}}      \\ 
			\midrule
			\cite{stradowskiMachineLearningSoftware2023a}            & 1,194                      & 742   & 451    & 62:38 & 0     & Information and Software Technology       & 4  & 4 & SE Process            \\ 
			\cite{alonsoSystematicMappingStudy2023b}                 & 223                        & 191   & 32     & 86:14 & 1     & Information and Software Technology       & 3  & 3 & SE Process                   \\ 
			\cite{somersDigitaltwinbasedTestingCyber2023b}           & 458                        & 146   & 312    & 32:68 & 53    & Information and Software Technology       & 3  & 4 & Software Testing                        \\ 
			\cite{sharbafConflictManagementTechniques2023a}          & 322                        & 171   & 151    & 53:47 & 0     & Software and Systems Modeling             & 3  & 6 & SE Models and Methods \\
			\cite{zakeri-nasrabadiSystematicLiteratureReview2023b}   & 10,454                     & 547   & 9,907  & 5:95  & 759   & Journal of Systems and Software           & 4  & 6 & Software Maintenance                   \\
			\cite{bjarnasonEmpiricallyBasedModel2023a}               & 4,671                      & 54    & 4,617  & 1:99  & 968   & Empirical Software Engineering            & 1  & 1 & Software Design                         \\
			\cite{lewowskiHowFarAre2022a}                            & 1,733                      & 169   & 1,564  & 10:90 & 82    & Information and Software Technology       & 13 & 0 & Software Testing                        \\
			\cite{vandinterPredictiveMaintenanceUsing2022a}          & 606                        & 144   & 462    & 24:76 & 47    & Information and Software Technology       & 3  & 3 & Software Maintenance / Operations       \\
			\cite{casalaroModeldrivenEngineeringMobile2022a}         & 2,541                      & 74    & 2,467  & 3:97  & 281   & Software and Systems Modeling             & 6  & 4 & Software Architecture                   \\
			\cite{deckersSystematicLiteratureReview2022a}            & 540                        & 36    & 504    & 7:93  & 219   & Journal of Systems and Software           & 3  & 6 & Software Requirements                   \\
			\cite{tambonHowCertifyMachine2022a}                      & 1,741                      & 290   & 1,451  & 17:83 & 717   & Automated Software Engineering            & 2  & 6 & Software Quality                        \\
			\cite{raniDecadeCodeComment2023a}                        & 2,353                      & 73    & 2,280  & 3:97  & 274   & Journal of Systems and Software           & 2  & 7 & Software Quality                        \\
			\cite{pereiraLearningSoftwareConfiguration2021a}         & 69                         & 69    & 0      & 100:0 & 0     & Journal of Systems and Software           & 3  & 4 & Software Configuration Management       \\
			\cite{rodriguez-perezPerceivedDiversitySoftware2021a}    & 3,194                      & 234   & 2,960  & 7:93  & 446   & Empirical Software Engineering            & 7  & 7 & SE Professional Practice                   \\
			\cite{barisicMultiparadigmModelingCyber2022a}            & 327                        & 153   & 174    & 47:53 & 1     & Journal of Systems and Software           & 5  & 6 & SE Models and Methods \\
			\cite{ebrahimiMobileAppPrivacy2021a}                     & 113                        & 55    & 58     & 49:51 & 2     & Information and Software Technology       & 3  & 3 & Software Quality                   \\
			\cite{tebesAnalyzingDocumentingSystematic2020a}          & 731                        & 522   & 209    & 72:28 & 236   & Information and Software Technology       & 3  & 7 & Software Testing                        \\
			\cite{geraldiSoftwareProductLine2020a}                   & 167                        & 56    & 111    & 34:66 & 0     & Information and Software Technology       & 6  & 8 & SE Models and Methods            \\
			\cite{lenarduzziSystematicLiteratureReview2021a}         & 318                        & 44    & 274    & 14:86 & 17    & Journal of Systems and Software           & 4  & 7 & Software Maintenance / SE Economics          \\
			\cite{kuutilaTimePressureSoftware2020b}                  & 5,454                      & 223   & 5,231  & 4:96  & 1,268 & Information and Software Technology       & 2  & 3 & SE Management / SE Professional Practice            \\
			\cite{shevtsovControlTheoreticalSoftwareAdaptation2018a} & 1,512                      & 161   & 1,351  & 11:89 & 25    & IEEE Transactions on Software Engineering & 4  & 0 & SE Models \& Methods \\
			\cite{teixeiraModelingAutomaticCode2017a}                & 318                        & 187   & 131    & 59:41 & 0     & Journal of Systems and Software           & 1  & 6 & SE Models \& Methods                   \\
			\cite{linOpinionMiningSoftware2022a}                     & 802                        & 127   & 675    & 16:84 & 21    & ACM Trans. Softw. Eng                     & 5  & 5 & SE Professional Practice                   \\
			\cite{zhangTestingVerificationNeuralnetworkbased2020a}   & 105                        & 105   & 0      & 100:0 & 1     & Information and Software Technology       & 3  & 4 & Software Testing                        \\
			\midrule
			\textbf{Total (24)}                                      & 39,946                     & 4,573 & 35,373 & 11:89 & 5,418 & \textbf{Median}                           & 3  & 4.5 &                                         \\
			\midrule
			\textbf{Dataset}                                           & \textbf{34,528} & \textbf{4,197} & \textbf{30,331} & \textbf{12:88} & \textbf{0} \\
			\bottomrule
		\end{tabular}
}%
\end{table*}

\label{section:results}

\subsection{\firstrq}

Yes, 24 secondary studies were found with title-abstract screening data. The process of how the dataset was created is explained in Section \ref{section:methods:datasetcreation}. In total, we reviewed 169 research artifacts, from which 11 (6.5\%) were inaccessible due to reasons such as dead links, empty repositories or pages requiring authorization. Ultimately, 24 of the 169 artifacts (14.2\%) were suitable for our benchmark dataset. The secondary studies selected for the benchmark dataset are shown in Table \ref{table:list_of_studies_with_screening_data}. The number of primary studies per secondary study varies significantly in the dataset — ranging from under a hundred papers to over 10,000 papers. Similar variations exist in the included/excluded ratio of papers. Two secondary studies contained only included studies, resulting in an I/E ratio of $100:0$, while at the other end of the spectrum, one study included only about one percent of papers, with an I/E ratio of $1:99$. On average, looking at the ratio of included and excluded studies in the dataset, 12\% of primary studies were labeled as included and 88\% as excluded. The median number of the inclusion and exclusion criteria across secondary studies was $3$ and $4.5$, respectively. Based on the Software Engineering Body of Knowledge (SWEBOK)~\cite{swebok}, the dataset covers a wide spectrum of SE domains.

Missing abstracts were an issue with many primary studies. Some abstracts were unavailable due to paywalls, and some primary studies were referenced only through citations. After automatically and manually retrieving abstracts for primary studies, as explained in Section \ref{section:methods:datasetcreation}, we were left with 5,418 (13.6\%) primary studies for which we could not retrieve abstracts. Primary studies with missing abstracts were removed from the final dataset. In total, the dataset contains 34,528 primary studies sourced from 24 secondary studies.

\subsection{\secondrq}

The results of LLM screening performance are included in Table \ref{table:llm_precision_recall_compact}. We report accuracy, precision, recall, and F1 score for both primary and secondary studies. This is because some secondary studies contain a large number of papers and therefore dominate the primary study scores. In contrast, averaging the secondary study scores gives equal weight to each secondary study. If you are a researcher wondering what average performance one might expect from LLMs in a secondary study, it is probably better to look at the secondary study lines.

The top seven tested LLMs showed similar levels of performance. Table \ref{table:llm_winners} lists the winning LLMs based on accuracy and F1 score in both primary and secondary studies. We can see that due to small differences between the top models, the best-performing models vary depending on whether we look at the F1 score, accuracy, or the primary and secondary categories. However, it is notable that the two smallest models — GPT-4.1 nano and Ministral 8B — exhibited poor performance in the screening task, indicating that smaller LLMs are not recommended.

\begin{table}[tbp]
    \centering
    \caption{Screening performance of LLMs across primary studies and secondary studies.}
    \label{table:llm_precision_recall_compact}
    \resizebox{\linewidth}{!}{%
    \begin{tabular}{lrlllll}
        \toprule
        \textbf{LLM} & \textbf{Average} & \textbf{Accuracy} & \textbf{Precision} & \textbf{Recall} & \multicolumn{2}{c}{\textbf{F1 score}} \\
        \midrule
        \multirow{2}{*}{\textbf{Llama 4 Maverick}}
        & Primary (n=34,528)     & 0.87 & 0.47 & 0.61 & \multicolumn{2}{c}{0.53} \\
        & Secondary (n=24)     & 0.74 & 0.54 & 0.61 & 0.50\footnotemark[1] & 0.57\footnotemark[2] \\
        \midrule
        \multirow{2}{*}{\textbf{o3-mini}}
        & Primary (n=34,528)     & 0.86 & 0.43 & 0.52 & \multicolumn{2}{c}{0.47} \\
        & Secondary (n=24)     & 0.73 & 0.60 & 0.49 & 0.46\footnotemark[1] & 0.54\footnotemark[2] \\
        \midrule
        \multirow{2}{*}{\textbf{GPT-4o}}
        & Primary (n=34,528)     & 0.84 & 0.40 & 0.66 & \multicolumn{2}{c}{0.50} \\
        & Secondary (n=24)     & 0.73 & 0.53 & 0.66 & 0.51\footnotemark[1] & 0.59\footnotemark[2] \\
        \midrule
        \multirow{2}{*}{\textbf{GPT-4.1}}
        & Primary (n=34,528)     & 0.83 & 0.38 & 0.63 & \multicolumn{2}{c}{0.47} \\
        & Secondary (n=24)     & 0.73 & 0.53 & 0.60 & 0.48\footnotemark[1] & 0.56\footnotemark[2] \\
        \midrule
        \multirow{2}{*}{\textbf{GPT-4.1 mini}}
        & Primary (n=34,528)     & 0.90 & 0.60 & 0.43 & \multicolumn{2}{c}{0.50} \\
        & Secondary (n=24)     & 0.73 & 0.62 & 0.38 & 0.41\footnotemark[1] & 0.47\footnotemark[2] \\
        \midrule
        \multirow{2}{*}{\textbf{GPT-4.1 nano}}
        & Primary (n=34,528)     & 0.56 & 0.19 & 0.81 & \multicolumn{2}{c}{0.31} \\
        & Secondary (n=24)     & 0.59 & 0.39 & 0.76 & 0.45\footnotemark[1] & 0.52\footnotemark[2] \\
        \midrule
        \multirow{2}{*}{\textbf{Claude 3.7 Sonnet}}
        & Primary (n=34,528)     & 0.89 & 0.56 & 0.46 & \multicolumn{2}{c}{0.51} \\
        & Secondary (n=24)     & 0.73 & 0.58 & 0.45 & 0.44\footnotemark[1] & 0.51\footnotemark[2] \\
        \midrule
        \multirow{2}{*}{\textbf{DeepSeek R1}}
        & Primary (n=34,528)     & 0.89 & 0.56 & 0.42 & \multicolumn{2}{c}{0.48} \\
        & Secondary (n=24)     & 0.72 & 0.58 & 0.41 & 0.41\footnotemark[1] & 0.48\footnotemark[2] \\
        \midrule
        \multirow{2}{*}{\textbf{Ministral 8B}}
        & Primary (n=34,528)     & 0.13 & 0.12 & 1.00 & \multicolumn{2}{c}{0.22} \\
        & Secondary (n=24)     & 0.34 & 0.34 & 1.00 & 0.44\footnotemark[1] & 0.51\footnotemark[2] \\
        \bottomrule
    \end{tabular}%
    }
\end{table}
\footnotetext[1]{Average F1 score of the F1 scores from the 24 secondary studies.}
\footnotetext[2]{F1 score calculated from the average precision and recall.}
\begin{table}[tbp]
    \centering
    \caption{Highest-scoring LLMs for different metrics and study types.}
    \label{table:llm_winners}
    \begin{tabular}{lll}
        \toprule
        \textbf{Metric} & \textbf{Study type}  & \textbf{Highest-scoring} \\
        \midrule
        Accuracy & Primary study  & GPT-4.1 mini \\
        Accuracy & Secondary study & Llama 4 Maverick \\
        F1 & Primary study & Llama 4 Maverick \\
        F1\footnotemark[1], F1\footnotemark[2] & Secondary study & GPT-4o \\
        \bottomrule
    \end{tabular}
\end{table}

\begin{figure}
\centering
\resizebox{0.45\columnwidth}{!}{%
\begin{tikzpicture}
\begin{axis}[
    xmin=1, xmax=7,
    ymin=0.0, ymax=1.0,
    xtick={1,2,3,4,5,6,7},
    xlabel={Likert-scale},
    ylabel={},
    grid=both,
    major grid style={line width=.2pt,draw=gray!50},
    minor tick num=1,
    enlargelimits=true,
    legend style={at={(0.6,0.25)},anchor=north west, nodes={scale=0.7, transform shape}}, 
    fill=none,
    legend columns=1
]
\addplot[
    thick,
    blue,
] coordinates {(1,0.122)
(2,0.141)
(3,0.241)
(4,0.269)
(5,0.269)
(6,0.269)
(7,0.498)};
\addplot[
    thick,
    orange
] coordinates  {(1,1.0)
(2,0.945)
(3,0.712)
(4,0.619)
(5,0.616)
(6,0.613)
(7,0.231)};
\addplot[
    thick,
    green
] coordinates {(1,0.217)
(2,0.245)
(3,0.36)
(4,0.375)
(5,0.375)
(6,0.374)
(7,0.316)};
\legend{Precision, Recall, F1}
\end{axis}
\end{tikzpicture}
}%
\caption{Average secondary study precision, recall and F1 for Likert-scale (1-7), across all LLMs.}
\label{figure:likert_avg_secondary_study_precision}
\end{figure}
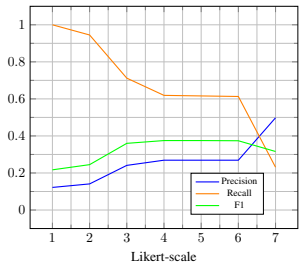

Table \ref{table:llm_precision_recall_compact} also reports precision and recall for the binary decisions made by the models. To highlight differences in precision and recall, we present Likert scale results in Fig.~\ref{figure:likert_avg_secondary_study_precision}, where the precision, recall, and F1 score curves at each decision point provide a clearer picture. The Likert scale is described in Section \ref{section:evaluation_llm_performance}. From Fig.~\ref{figure:likert_avg_secondary_study_precision} we observe that the F1 score behaves as expected, forming an inverted U-shape and peaking in the middle of the Likert scale. Recall decreases as we move along the X-axis from left to right, dropping from a perfect $1.00$ to $0.23$. Precision exhibits the opposite trend, increasing from $0.12$ to $0.50$.

Often, a researcher using the LLMs wants to optimize for maximum recall, as missing evidence (false negatives) is more difficult to recover from than incorrectly including a paper (false positive). Here, we assume a process where an LLM performs the initial title-abstract screening, and then the included papers are checked by the researcher. In such a setup, a deemed useful LLM could be defined as follows. We would aim for high recall, at least above $0.95$, with reasonable precision, around $0.50$. This would mean capturing 95\% of the evidence, while manually screening papers of which 50\% contain relevant evidence. Unfortunately, as seen in Fig. \ref{figure:likert_avg_secondary_study_precision}, such a point is not reached with current LLMs.

\subsection{\thirdrq}

Table \ref{table:secondary_study_precision_recall_compact} shows the accuracy, precision, recall, and F1 score for the secondary studies across all LLMs. We observe notable variation between the secondary studies, with accuracy ranging from $0.34$ to $0.85$ and F1 scores ranging from $0.07$ to $0.92$. In no study do we reach our deemed useful thresholds for LLMs, as none achieve a recall above 0.95 while maintaining a precision of $0.50$.

\renewcommand{\thesubfigure}{\roman{subfigure}}
\begin{figure*}[tbp]
\centering
\newlength{\figwidth}
\setlength{\figwidth}{0.194\textwidth}
\begin{subfigure}[]{\figwidth}
    \subcaption{\cite{alonsoSystematicMappingStudy2023b}}
    \resizebox{\columnwidth}{!}{%
    \begin{tikzpicture}
    \begin{axis}[
        xmin=1, xmax=7,
        ymin=0.0, ymax=1.0,
        xtick={1,2,3,4,5,6,7},
        xlabel={},
        ylabel={},
        grid=both,
        major grid style={line width=.2pt,draw=gray!50},
        minor tick num=1,
        enlargelimits=true,
        legend style={at={(0.06,0.25)},anchor=north west, nodes={scale=0.7, transform shape}}, 
        fill=none,
        legend columns=1
    ]
    \addplot[
        thick,
        blue,
    ] coordinates {(1,0.856)
(2,0.817)
(3,0.805)
(4,0.832)
(5,0.832)
(6,0.831)
(7,0.841)};
    \addplot[
        thick,
        orange
    ] coordinates  {(1,1.0)
(2,0.674)
(3,0.344)
(4,0.286)
(5,0.284)
(6,0.276)
(7,0.034)};
    \addplot[
        thick,
        green
    ] coordinates {(1,0.922)
(2,0.738)
(3,0.482)
(4,0.426)
(5,0.423)
(6,0.414)
(7,0.065)};
    \legend{Precision, Recall, F1}
    \end{axis}
    \end{tikzpicture}
    }%
    \end{subfigure}
\begin{subfigure}[]{\figwidth}
    \subcaption{\cite{sharbafConflictManagementTechniques2023a}}
    \resizebox{\columnwidth}{!}{%
    \begin{tikzpicture}
    \begin{axis}[
        xmin=1, xmax=7,
        ymin=0.0, ymax=1.0,
        xtick={1,2,3,4,5,6,7},
        xlabel={},
        ylabel={},
        grid=both,
        major grid style={line width=.2pt,draw=gray!50},
        minor tick num=1,
        enlargelimits=true,
        legend style={at={(0.06,0.25)},anchor=north west, nodes={scale=0.7, transform shape}}, 
        fill=none,
        legend columns=1
    ]
    \addplot[
        thick,
        blue,
    ] coordinates {(1,0.531)
(2,0.535)
(3,0.606)
(4,0.656)
(5,0.657)
(6,0.658)
(7,0.802)};
    \addplot[
        thick,
        orange
    ] coordinates  {(1,1.0)
(2,0.988)
(3,0.833)
(4,0.734)
(5,0.734)
(6,0.73)
(7,0.277)};
    \addplot[
        thick,
        green
    ] coordinates {(1,0.694)
(2,0.694)
(3,0.701)
(4,0.693)
(5,0.693)
(6,0.692)
(7,0.412)};
    \legend{Precision, Recall, F1}
    \end{axis}
    \end{tikzpicture}
    }%
    \end{subfigure}
\begin{subfigure}[]{\figwidth}
    \subcaption{\cite{pereiraLearningSoftwareConfiguration2021a}}
    \resizebox{\columnwidth}{!}{%
    \begin{tikzpicture}
    \begin{axis}[
        xmin=1, xmax=7,
        ymin=0.0, ymax=1.0,
        xtick={1,2,3,4,5,6,7},
        xlabel={},
        ylabel={},
        grid=both,
        major grid style={line width=.2pt,draw=gray!50},
        minor tick num=1,
        enlargelimits=true,
        legend style={at={(0.06,0.25)},anchor=north west, nodes={scale=0.7, transform shape}}, 
        fill=none,
        legend columns=1
    ]
    \addplot[
        thick,
        blue,
    ] coordinates {(1,1.0)
(2,1.0)
(3,1.0)
(4,1.0)
(5,1.0)
(6,1.0)
(7,1.0)};
    \addplot[
        thick,
        orange
    ] coordinates  {(1,1.0)
(2,0.994)
(3,0.887)
(4,0.847)
(5,0.847)
(6,0.847)
(7,0.409)};
    \addplot[
        thick,
        green
    ] coordinates {(1,1.0)
(2,0.997)
(3,0.94)
(4,0.917)
(5,0.917)
(6,0.917)
(7,0.581)};
    \legend{Precision, Recall, F1}
    \end{axis}
    \end{tikzpicture}
    }%
    \end{subfigure}
\begin{subfigure}[]{\figwidth}
    \subcaption{\cite{tebesAnalyzingDocumentingSystematic2020a}}
    \resizebox{\columnwidth}{!}{%
    \begin{tikzpicture}
    \begin{axis}[
        xmin=1, xmax=7,
        ymin=0.0, ymax=1.0,
        xtick={1,2,3,4,5,6,7},
        xlabel={},
        ylabel={},
        grid=both,
        major grid style={line width=.2pt,draw=gray!50},
        minor tick num=1,
        enlargelimits=true,
        legend style={at={(0.06,0.25)},anchor=north west, nodes={scale=0.7, transform shape}}, 
        fill=none,
        legend columns=1
    ]
    \addplot[
        thick,
        blue,
    ] coordinates {(1,0.721)
(2,0.748)
(3,0.776)
(4,0.761)
(5,0.76)
(6,0.76)
(7,0.825)};
    \addplot[
        thick,
        orange
    ] coordinates  {(1,1.0)
(2,0.883)
(3,0.426)
(4,0.322)
(5,0.322)
(6,0.322)
(7,0.059)};
    \addplot[
        thick,
        green
    ] coordinates {(1,0.838)
(2,0.81)
(3,0.55)
(4,0.453)
(5,0.453)
(6,0.453)
(7,0.11)};
    \legend{Precision, Recall, F1}
    \end{axis}
    \end{tikzpicture}
    }%
    \end{subfigure}
\begin{subfigure}[]{\figwidth}
    \subcaption{\cite{teixeiraModelingAutomaticCode2017a}}
    \resizebox{\columnwidth}{!}{%
    \begin{tikzpicture}
    \begin{axis}[
        xmin=1, xmax=7,
        ymin=0.0, ymax=1.0,
        xtick={1,2,3,4,5,6,7},
        xlabel={},
        ylabel={},
        grid=both,
        major grid style={line width=.2pt,draw=gray!50},
        minor tick num=1,
        enlargelimits=true,
        legend style={at={(0.06,0.25)},anchor=north west, nodes={scale=0.7, transform shape}}, 
        fill=none,
        legend columns=1
    ]
    \addplot[
        thick,
        blue,
    ] coordinates {(1,0.588)
(2,0.611)
(3,0.703)
(4,0.716)
(5,0.716)
(6,0.721)
(7,0.878)};
    \addplot[
        thick,
        orange
    ] coordinates  {(1,1.0)
(2,0.913)
(3,0.519)
(4,0.419)
(5,0.419)
(6,0.417)
(7,0.068)};
    \addplot[
        thick,
        green
    ] coordinates {(1,0.741)
(2,0.732)
(3,0.597)
(4,0.528)
(5,0.528)
(6,0.528)
(7,0.127)};
    \legend{Precision, Recall, F1}
    \end{axis}
    \end{tikzpicture}
    }%
\end{subfigure}
\caption{Average precision, recall and F1 for Likert scale (1-7), across all LLMs for secondary studies with vastly different curve shape compared to Fig. \ref{figure:likert_avg_secondary_study_precision}.}
\label{figure:likert_results}
\end{figure*}
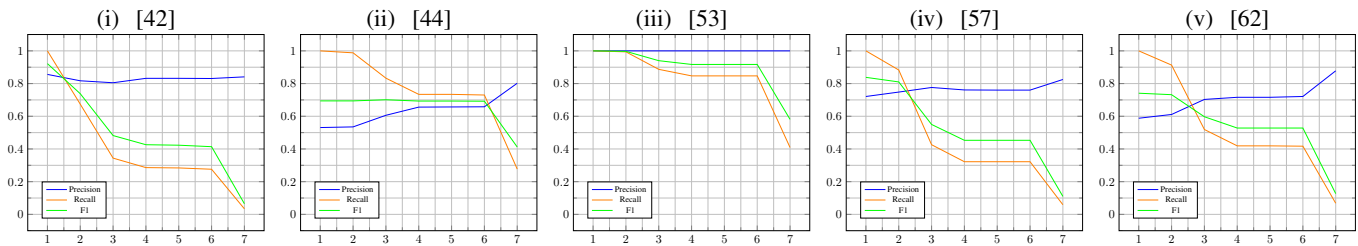

Fig. \ref{figure:likert_results} shows the precision and recall curves for secondary studies that have a vastly different curve shape than in Fig. \ref{figure:likert_avg_secondary_study_precision}. We notice that for the study \cite{pereiraLearningSoftwareConfiguration2021a}, there is actually a point on the curve where the deemed useful threshold is reached. However, it is important to note that this is likely because the dataset for that paper contains only included studies, resulting in perfect precision and recall when all papers are included at the lowest Likert scores. If we consider this study an outlier, we are left with no cases where the deemed useful thresholds are met.

Looking at the F1 curves also reveals variability between studies. Most studies have the highest F1 scores in the middle of the Likert scale, as one would expect. However, there are multiple studies where the F1 score drops when moving from left to right along the X-axis \cite{alonsoSystematicMappingStudy2023b, sharbafConflictManagementTechniques2023a, pereiraLearningSoftwareConfiguration2021a, tebesAnalyzingDocumentingSystematic2020a, teixeiraModelingAutomaticCode2017a}. These studies would clearly benefit from using the lowest values on the Likert scale and suffer from the binary decision threshold used in our experiment, which mostly corresponds to the median value on the Likert scale—number 4 in the figure.

\begin{table}[tbp]
    \centering
    \caption{Accuracy, precision, recall, and F1 score
for the secondary studies across all LLMs.}
    \begin{tabular}{lllll}
        \toprule
        \textbf{\makecell[l]{Study}} & \textbf{Accuracy} & \textbf{Precision} & \textbf{Recall} & \textbf{F1 score} \\
        \midrule
        \cite{stradowskiMachineLearningSoftware2023a} & 0.78 & 0.80 & 0.86 & 0.83 \\
        \cite{alonsoSystematicMappingStudy2023b} & 0.34 & 0.83 & 0.28 & 0.42 \\
        \cite{somersDigitaltwinbasedTestingCyber2023b} & 0.66 & 0.49 & 0.55 & 0.51 \\
        \cite{sharbafConflictManagementTechniques2023a} & 0.66 & 0.66 & 0.73 & 0.69 \\
        \cite{zakeri-nasrabadiSystematicLiteratureReview2023b} & 0.72 & 0.13 & 0.76 & 0.22 \\
        \cite{bjarnasonEmpiricallyBasedModel2023a} & 0.80 & 0.03 & 0.57 & 0.07 \\
        \cite{lewowskiHowFarAre2022a} & 0.78 & 0.28 & 0.75 & 0.41 \\
        \cite{vandinterPredictiveMaintenanceUsing2022a} & 0.63 & 0.39 & 0.80 & 0.52 \\
        \cite{casalaroModeldrivenEngineeringMobile2022a} & 0.79 & 0.10 & 0.78 & 0.18 \\
        \cite{deckersSystematicLiteratureReview2022a} & 0.67 & 0.14 & 0.82 & 0.23 \\
        \cite{tambonHowCertifyMachine2022a} & 0.76 & 0.39 & 0.55 & 0.45 \\
        \cite{raniDecadeCodeComment2023a} & 0.83 & 0.08 & 0.40 & 0.14 \\
        \cite{pereiraLearningSoftwareConfiguration2021a} & 0.85 & 1.00 & 0.85 & 0.92 \\
        \cite{rodriguez-perezPerceivedDiversitySoftware2021a} & 0.80 & 0.16 & 0.35 & 0.22 \\
        \cite{barisicMultiparadigmModelingCyber2022a} & 0.64 & 0.65 & 0.50 & 0.57 \\
        \cite{ebrahimiMobileAppPrivacy2021a} & 0.59 & 0.56 & 0.82 & 0.66 \\
        \cite{tebesAnalyzingDocumentingSystematic2020a} & 0.44 & 0.76 & 0.32 & 0.45 \\
        \cite{geraldiSoftwareProductLine2020a} & 0.59 & 0.42 & 0.57 & 0.48 \\
        \cite{lenarduzziSystematicLiteratureReview2021a} & 0.58 & 0.14 & 0.36 & 0.20 \\
        \cite{kuutilaTimePressureSoftware2020b} & 0.82 & 0.13 & 0.46 & 0.21 \\
        \cite{shevtsovControlTheoreticalSoftwareAdaptation2018a} & 0.71 & 0.22 & 0.64 & 0.32 \\
        \cite{teixeiraModelingAutomaticCode2017a} & 0.56 & 0.72 & 0.42 & 0.53 \\
        \cite{linOpinionMiningSoftware2022a} & 0.75 & 0.37 & 0.76 & 0.50 \\
\cite{zhangTestingVerificationNeuralnetworkbased2020a} & 0.40 & 1.00 & 0.40 & 0.57 \\
        \bottomrule
    \end{tabular}%
    \label{table:secondary_study_precision_recall_compact}
\end{table}

\subsection{\fourthrq}
\begin{table}[tbp]
\caption{Logistic regression model predicting the correctness of a single screening decision with the LLM and the secondary study.} 
\label{table:logistic_regression}
\resizebox{\linewidth}{!}{%
\begin{tabular}{llccccc}
\toprule
  & \textbf{Est.} & \textbf{std.err} & \textbf{$t$} & \textbf{$p$} & \makecell{\textbf{Odds}\\\textbf{ratio}}   \\
\midrule 
(Intercept)                            &           3.18 &  0.14  & 22.23  &  \textless{} 2e-16 *** &  24.08 \\
\midrule 
\addlinespace
\multicolumn{7}{l}{\textbf{Reference -   GPT-4.1 mini}}                                     \\
\addlinespace
Claude 3.7 Sonnet        &           -0.066   &   0.03   &  -2.62  &  0.008806 **   & 0.94 \\
DeepSeek R1           &            -0.076  &  0.025 &  -2.99 & 0.002755 **  &  0.93\\
GPT-4.1                 &        -0.612   &  0.023  &  -26.20  & \textless{} 2e-16 *** & 0.54 \\
GPT-4.1 nano           &        -2.053 &   0.021 & -95.83 &  \textless{} 2e-16 ***  & 0.13 \\
GPT-4o               &           -0.547   & 0.024 & -23.21 & \textless{} 2e-16 *** & 0.58 \\
Llama 4 Maverick           &      -0.295  &   0.024 &  -12.09 &  \textless{} 2e-16 *** & 0.74\\
Ministral 8B              &           -4.332  &   0.025 &  -174.75 &  \textless{} 2e-16 ***  & 0.01\\
o3-mini                &         -0.370  &   0.024 &  -15.33 &  \textless{} 2e-16 ***  & 0.69 \\
\hdashline
\addlinespace
\multicolumn{7}{l}{\textbf{Reference - \cite{pereiraLearningSoftwareConfiguration2021a}}} \\
\addlinespace
\cite{stradowskiMachineLearningSoftware2023a} & -0.702  &  0.145 &  -4.85 & \textless{}  2e-16 ***& 0.50\\
\cite{alonsoSystematicMappingStudy2023b} & -3.240  &  0.151 & -21.50 & \textless{}  2e-16 *** & 0.04\\
\cite{somersDigitaltwinbasedTestingCyber2023b} & -1.599  &  0.148 & -10.82 & \textless{}  2e-16 *** & 0.20\\
\cite{sharbafConflictManagementTechniques2023a} & -1.621   &  0.149 &  -10.87 &  \textless{}  2e-16 *** & 0.20 \\
\cite{zakeri-nasrabadiSystematicLiteratureReview2023b}  &  -1.194  &  0.142  & -8.39 & \textless{} 2e-16 *** & 0.30\\
\cite{bjarnasonEmpiricallyBasedModel2023a} & -0.493  &  0.143 &  -3.45 & 0.000570 ***& 0.61 \\
\cite{lewowskiHowFarAre2022a} & -0.697  &   0.144 & -4.84 & 1.33e-06 *** & 0.50 \\
\cite{vandinterPredictiveMaintenanceUsing2022a} & -1.775  &  0.146 & -12.16 & \textless{} 2e-16 *** & 0.17\\
\cite{casalaroModeldrivenEngineeringMobile2022a}  & -0.665  &  0.143  & -4.63  & 3.66e-06 *** & 0.51\\
\cite{deckersSystematicLiteratureReview2022a}  & -1.554  &  0.149 & -10.40 & \textless{} 2e-16 ***& 0.21\\
\cite{tambonHowCertifyMachine2022a}&  -0.920  &    0.145   &  -6.29  &  3.12e-10 ***  & 0.40 \\
\cite{raniDecadeCodeComment2023a}  & -0.183  &  0.144  & -1.27 & 0.20  & 0.83 \\
\cite{rodriguez-perezPerceivedDiversitySoftware2021a} &  -0.510  &  0.143 & -3.56  & 0.000376 *** & 0.60 \\
\cite{barisicMultiparadigmModelingCyber2022a}  & -1.720 &   0.149 & -11.55 & \textless{} 2e-16 *** & 0.18 \\
\cite{ebrahimiMobileAppPrivacy2021a} & -2.017 &  0.160 &  -12.63 &  \textless{}  2e-16 *** & 0.13 \\
\cite{tebesAnalyzingDocumentingSystematic2020a}  & -2.760  &  0.146 & -18.91 & \textless{}  2e-16 ***& 0.06\\
\cite{geraldiSoftwareProductLine2020a} & -1.989   & 0.154 & -12.91 & \textless{}  2e-16 ***& 0.14\\
\cite{lenarduzziSystematicLiteratureReview2021a}  & -2.024 &  0.149 & -13.60 & \textless{} 2e-16 *** & 0.13\\
\cite{kuutilaTimePressureSoftware2020b} & -0.271  &  0.143 &  -1.89 & 0.06 &  0.76 \\
\cite{shevtsovControlTheoreticalSoftwareAdaptation2018a}  & -1.264   &  0.144  &  -8.79  & \textless{}  2e-16 *** & 0.28 \\
\cite{teixeiraModelingAutomaticCode2017a}  & -2.144  &  0.148 & -14.46 & \textless{} 2e-16 *** & 0.12\\
\cite{linOpinionMiningSoftware2022a} & -0.931  &  0.146 &  -6.38 & 1.81e-10 ***&  0.39\\
\cite{zhangTestingVerificationNeuralnetworkbased2020a} &-2.942  &  0.159 & -18.48 & \textless{} 2e-16 *** & 0.05 \\
\bottomrule 
\end{tabular}
}%
\end{table}
As both the LLMs and secondary studies impact the results of LLM-based paper screening, we decided to investigate how these factors work together. We did this by fitting a logistic regression model predicting the correctness of screening decisions based on the LLM used and the secondary study. We selected the best-performing model and secondary study for reference categories, as this makes the resulting model easier to interpret. The model is shown in Table~\ref{table:logistic_regression}. 
In the model, a statistically significant intercept (log-odds = $3.20$, p \textless{} $2\mathrm{e}{-16}$) corresponds to high baseline odds of a correct prediction when using the reference model (GPT-4.1 mini) and reference study (\cite{pereiraLearningSoftwareConfiguration2021a}) (odds ratio $\approx 24.08$).

As the best performing LLM was selected to be the reference category, all other LLMs have reduced odds of correctness relative to the baseline. Both Ministral 8B and GPT-4.1 nano showed a notably large negative effect ($\beta$ = $-4.332$, p \textless{} $2\mathrm{e}{-16}$, OR $\approx 0.013$) ($\beta$ = $-2.05$, p \textless{} $2\mathrm{e}{-16}$, OR $\approx 0.128$), suggesting substantially poorer performance compared to the larger LLMs that had notable better coefficient and odds-rations.  

Other LLMs such as GPT-4o (OR $\approx 0.58$) and GPT-4.1 (OR $\approx 0.54$) also demonstrated significantly diminished performance. However, LLMs such as Claude 3.7 Sonnet and DeepSeek R1 were close to the performance of the reference LLM (ORs $\approx 0.93-0.94$).

Regarding the influence of secondary studies, the regression model identified considerable variability in the odds of correct classification across sources. Several secondary studies were associated with large negative effects. For example, screening tasks from Alonso et al.~\cite{alonsoSystematicMappingStudy2023b} and Tebes et al.\cite{tebesAnalyzingDocumentingSystematic2020a} exhibited strong reductions in predictive accuracy (ORs $\approx 0.05$ and $0.06$, respectively). Conversely, a few studies (e.g., Rani et al.~\cite{raniDecadeCodeComment2023a} and Kuutila et al.\cite{kuutilaTimePressureSoftware2020b}, OR $\approx 0.83$ and $0.78$) had relatively weaker effects. Thus, both the LLM used and the characteristics of the secondary study significantly impact the likelihood of correct screening in our dataset, when using zero-shot prompts.

\subsection{\fifthrq}

We benchmarked 9 different LLMs across 34,528 primary studies. The total number of screening decisions we ran with the LLMs was 302,787 (9 * 34,528). This allowed the precise measurement of LLM token usage and time per paper -- metrics, which can be useful for other researchers and for future cost approximation, as LLMs are typically billed at the token level. Table \ref{table:llm_cost} summarizes the tested LLMs with their corresponding token costs, average token usage per secondary study, the 95th percentile time per primary study and the cost per secondary study.

In terms of time efficiency, we found that GPT-4.1 mini and GPT-4o achieved performance comparable with slower models (see Table \ref{table:llm_precision_recall_compact}), while offering a three- to five-fold reduction in screening time per paper compared to the highest-performing models (see Table \ref{table:llm_cost}). Among the three fastest models, Ministral 8B and GPT-4.1 nano -- screening performance was notably limited. For comparison, Huotala et al.~\cite{huotalaPromiseChallengesUsing2024a} reported that expert human screeners required an average of 85.95 seconds to screen a single paper. In contrast, every LLM except o3-mini were faster than the human screener.

Table \ref{table:llm_cost} also presents the average number of tokens required to evaluate a secondary study across models. The input token count was generally consistent between LLMs, although some models hosted via OpenRouter required additional prompt instructions to properly output JSON. The most token-efficient LLM we tested was GPT-4.1 mini, which delivered performance on par with o3-mini while using ten times as less output tokens (Table \ref{table:llm_precision_recall_compact}). With the reasoning models, we noticed a 10-fold increase in output tokens of o3-mini, which is mainly because the reasoning LLMs output "reasoning tokens" to aid in its decision chain. We didn't observe the same for the second reasoning model (DeepSeek R1), as the OpenRouter API did not consistently return the reasoning token count.

Regarding costs, we observe that even the most expensive model (o3-mini) cost only $24 * \$36.7\approx \$881$ across the 24 secondary studies. As such, it appears that the cost of using LLMs should not be a barrier to their use in title-abstract screening. When it comes to cost-efficiency, we find that Meta's LLaMA 4 Maverick (cost per secondary study \$1.4) and GPT-4.1 mini (cost per secondary study \$2.2) delivered performance on par with more expensive models (see Table \ref{table:llm_precision_recall_compact}), while offering over tenfold savings in cost compared to the more expensive models (see Table \ref{table:llm_cost}). 

\begin{table}[tbp]
    \centering
    \caption{Token costs ($^{*}$as of April 2025) of the LLMs, average tokens used per secondary study, the time per paper and cost per secondary study when running the experiments.}
    \label{table:llm_cost}
\resizebox{\linewidth}{!}{
    \begin{tabular}{lcccccc}
    \toprule
    \textbf{LLM} & \makecell[c]{\textbf{Input} \\ \textbf{\$ / 1M} \\ \textbf{Tokens\small $^{*}$}} & \makecell[c]{\textbf{Output} \\ \textbf{\$ / 1M} \\ \textbf{Tokens\small $^{*}$}} & \textbf{\makecell[c]{Input\\tokens\\(per study)}} & \textbf{\makecell[c]{Output\\tokens\\(per study)}} & \textbf{\makecell[c]{Time (P95) \\Per paper}} & \makecell[r]{\textbf{Cost per} \\ \textbf{secondary} \\ \textbf{study\small $^{*}$}} \\
    \midrule
    o3-mini  & \$1.10 & \$4.40 & 2.43M & 7.73M & 86.1s & \$36.7  \\
    GPT-4o & \$2.50  & \$10.00 & 2.43M & 0.73M  & 14.1s & \$13.4  \\
    GPT-4.1 & \$2.00  & \$8.00 & 2.43M & 0.89M & 24.7s & \$12.1  \\
    GPT-4.1 mini & \$0.40  & \$1.60 & 2.43M & 0.77M & 10.5s & \$2.2  \\
    GPT-4.1 nano & \$0.10  & \$0.40 & 2.43M & 0.72M & 7.2s & \$0.5  \\
    Claude 3.7 Sonnet & \$3.00 & \$10.00 & 3.37M & 1.39M & 33.7s & \$24.0  \\
    \midrule
    DeepSeek R1 & \$0.54 & \$2.18 & 2.82M & 1.03M & 42.8s & \$3.8  \\
    Llama 4 Maverick & \$0.19& \$0.85 & 2.86M & 1.06M & 23.0s & \$1.4 \\
    Ministral 8B & \$0.10 & \$0.10 & 2.89M & 1.03M & 12.0s & \$0.4 \\
    \bottomrule
    \end{tabular}
}%
\end{table}

\section{Discussion}
\label{section:discussion}

\subsection{Creating the benchmark dataset}

This paper presents one of the largest title-abstract screening benchmarks. Comparison to prior works (Table \ref{table:priorworks}) shows that in terms of the number of secondary studies, SESR-Eval is the largest (24 secondary studies). We have the second highest number of LLMs benchmarked (9 LLMs) and the second highest number of primary studies (34,528).

One may ask why the number of secondary studies is relatively low in the dataset, and will it increase rapidly in the future. Based on our experience, we argue that such an increase is unlikely. The creation of our benchmark relied on the quality of the research artifacts of the secondary studies. However, the lack of standardized practices for reporting research artifacts presents significant challenges.

We faced multiple challenges due to the structure and representation of the data and the research artifacts. Although most of the research artifacts included only spreadsheet files (CSV and Excel), some packages included files that we could not open due to proprietary file format. This highlights the need for using standard file formats in research artifacts, as not all researchers own licenses for proprietary or deprecated software. Each research artifact is unique, requiring effort to understand. Examining these artifacts resembles manual reverse-engineering, where the researcher must reconstruct the process behind the original analysis. For instance, inconsistent column formats posed a challenge for retrieving the ground truth, as the decision was fragmented into multiple columns. In addition, included and excluded papers were often distributed across multiple files, which required manual verification. Finaly, converting each research artifact into a unified format suitable for benchmarking purposes must be done case-by-case bases.

Some prior studies~\cite{kusa2023csmed} have bypassed the challenges associated with research artifacts. They reused the original authors' search queries, re-executed them, and then treated the studies explicitly included in the published review as the included set, with the remainder of the query results inferred as excluded studies. While this approach simplifies the data collection, it does not accurately replicate the original review process. It fails to capture the accurate set of studies that were excluded during the review. It also does not capture the phase in which a paper was excluded, e.g., a paper might be included based on title-abstract screening and later excluded in the full-paper screening phase. Finally, search queries are difficult to replicate retrospectively, as academic databases are continuously updated. Thus, this method introduces uncertainty in identifying excluded studies.

\subsection{Screening performance - LLMs vs secondary studies}

While the seven best performing LLMs we tested offered similar performance, no LLM reached a high recall while maintaining reasonable precision. We suspect this is a limitation of the current LLM screening architecture. The two smallest models (GPT-4.1 nano and Ministral 8B) underperformed relative to the larger LLMs. This suggests that, once a suitable model size is reached, the specific choice of LLM is less critical than the fit between LLM-based screening and the characteristics of the secondary study being screened. A somewhat similar finding has been reported in the biomedical domain in the study by Dennstadt~\cite{dennstadtTitleAbstractScreening2024}. They screened 38,426 primary studies from 10 secondary studies and observed that performance varied depending on both the LLM and the secondary study used, which aligns with our observations. 

Dennstadt~\cite{dennstadtTitleAbstractScreening2024} also tested the use of a Likert scale in screening, which had a notable impact on performance. We similarly observed that choosing a different Likert scale point for the inclusion boundary had a notable effect and introduced variation between studies.

Work by Delgado~\cite{delgado-chavesTransformingLiteratureScreening2025} represents the largest prior study in terms of the number of LLMs evaluated (18), although they included only three secondary studies. In the medical domain, they reported screening accuracies of $0.92$, $0.88$, and $0.40$ for their three secondary studies. Our results are closely aligned with Delgado's, as the screening accuracy across our secondary studies ranged from $0.34$ to $0.75$ (Table $\ref{table:secondary_study_precision_recall_compact}$).

The related work in SE~\cite{huotalaPromiseChallengesUsing2024a, felizardoChatGPTApplicationSystematic2024} reported similar performance for LLMs as we did. However, they only evaluated a single LLM with two studies~\cite{felizardoChatGPTApplicationSystematic2024}, and two LLMs with a single SR~\cite{huotalaPromiseChallengesUsing2024a}. Thus, our numbers in terms of SRs and LLMs far exceed those. The first work~\cite{huotalaPromiseChallengesUsing2024a} reported a precision of $0.50$ and recall of $0.42$ when screening 1,306 primary studies from a single secondary study using GPT-4. Compared to our results, we surpass both the precision and recall with Claude 3.7 Sonnet ($0.56$ and $0.46$), while GPT-4.1 mini offers the best precision in our study ($0.60$ and $0.43$ for precision and recall). The second work~\cite{felizardoChatGPTApplicationSystematic2024} used one LLM and two primary studies and reported accuracies between $0.63$ and $0.86$. Our study exceeds that range, which is not surprising as we had more LLMs and more secondary studies under evaluation.

All prior works are listed in Table \ref{table:priorworks}. To summarize, our findings combined with the related work highlight that (1) Likert scale has a significant difference if one wants to maximize recall and/or precision; (2) The choice of the secondary study has a more significance than the LLM; and (3) Newer and smaller LLMs are on the same level as older, larger LLMs.

\subsection{Combined effect of LLMs and secondary studies}

Regression analysis coefficients highlighted that both the secondary study and the LLM have a highly significant impact on screening performance. It is notable that after we exclude the two smallest and cheapest LLMs, the difference in performance between the remaining LLMs is much smaller than the difference in performance between secondary studies, in terms of regression coefficients or odds ratios. So the takeaway message is: choose any large enough LLM, and the performance is similar, but one is still left with the between-study variance. We are not aware of prior studies that have statistically examined the combined effect of the secondary study and LLMs.

\subsection{Time, token efficiency, and cost of the LLMs}

From the tested LLMs and their corresponding timings (Table \ref{table:llm_cost}), we see that the fastest LLM per paper was GPT-4.1 nano. However, it was the second-worst performing LLM in the screening tasks. This means that it is not advisable to choose the fastest model for title-abstract screening. On the other hand, looking at the highest-scoring LLMs from Table \ref{table:llm_winners} and their corresponding screening performance in Table \ref{table:llm_precision_recall_compact}, Llama 4 Maverick is the highest scoring LLM in two categories: (1) Secondary study accuracy and (2) Primary study F1 score. Looking at the costs, Llama 4 Maverick cost only \$1.9 per secondary study, which is many times cheaper than most of the tested LLMs. This highlights that it is not optimal to always choose the most expensive LLM for title-abstract screening. Yet, it takes 23 seconds for Llama 4 Maverick to screen one primary study, which was much slower than most of the tested LLMs. However, requests to LLMs can be parallelized, which reduces the total time to screen the primary studies substantially. For a balanced selection, it would make sense to choose an LLM that screens papers efficiently; meaning that it screens papers quickly with an acceptable accuracy; and has reasonable costs.

\subsection{Limitations}
\label{section:limitations}
\noindent Next we outline the main limitations of our study. 

\subsubsection{Missing titles and abstracts} After missing abstract retrieval, the secondary studies we evaluated contained 5,418 missing abstracts, i.e., 13.6\% of all primary studies. Missing abstracts might have an effect on the results of our study.  As title-abstract screening requires a title and abstract to be present, removing the missing titles and abstracts was a required action for our study to be accurately benchmarking of LLMs in the title-abstract screening process. 

\subsubsection{Extraction of ground truth} Ground truth for the primary studies were extracted manually from each secondary study, using the research artifact's data fields and by full-text screening of the secondary study. Contacting the authors of the secondary studies about the ground truth could have reduced the likelihood of errors in this manual step.

\subsubsection{Only zero-shot prompting technique tested} As our dataset has over 30,000 primary studies while targeting to cover as much secondary studies as possible, we only ran prompts that do not include task examples (zero-shot), or additional reasoning instructions (chain-of-thought). Although previous studies have experimented on various different prompting techniques (e.g. one-shot, few-shot, chain-of-thought)~\cite{huotalaPromiseChallengesUsing2024a}, we believe that the reasoning LLMs, which in our case were o3-mini and DeepSeek R1, are sufficient to overcome this limitation. Testing different prompting techniques is a feasible area for future research that our dataset enables.

\subsubsection{Possible LLM bias and hallucinations} It is known that LLMs tend to hallucinate and that testing only single-vendor LLMs or a small subset of LLMs can introduce bias. To mitigate these threats, we tested 9 different LLMs across different vendors - with both reasoning, non-reasoning models and open-source models. Testing additional LLMs can be accomplished with minimal effort using our research artifact. In addition to these threats, using the same prompt template for all LLMs may introduce bias. To ensure a fair and direct comparison between LLMs, we believe using the same prompt template across LLMs is justified.

\subsubsection{Data extraction, parsing and transformation}
As discussed, the data extraction, parsing and transformation processes were time and labor-intensive, done on a case-by-case basis. The process had many parts, which introduced the potential for human error. For instance, the research artifact's files could be misunderstood. The chance of misunderstanding is greatly reduced if the research artifact contained good documentation. Although the authors made great effort to ensure accuracy, mistakes still may have occurred during the manual steps.

\subsubsection{Reliability of screening results} To obtain screening results for each of the secondary studies, we had to analyze the contents of the research artifacts and full-text screen secondary studies. The authors conducted inter-rater reliability assessments of the selected secondary studies to minimize the risk of including packages that are not suitable for the dataset. However, there is a risk that the inclusion / exclusion criteria in the secondary studies may include criterion, that cannot be evaluated solely based on title and abstract. We opted not to remove the criteria that cannot be purely evaluated by title and abstract, because this would introduce bias and potentially falsify research results.

\section{Conclusions}
\label{section:conclusions}

In this study we created the SESR-Eval dataset, which enables monitoring AI performance in the screening task of SRs in SE secondary studies. The dataset contains 34,528 labeled primary studies from 24 secondary studies. We benchmarked the title-abstract screening performance of 9 different LLMs with this dataset. So far, LLMs are not yet recommended for automating the screening process, as the accuracy varies widely across secondary studies and no LLM managed a high recall with reasonable precision. We found that choosing the most efficient LLM for the screening task is important, as it gives the optimal balance of speed, screening performance and cost. Finally, it appears that costs should not be a barrier in adapting LLMs for title-abstract screening, as even the most expensive model cost less than \$40 per secondary study.
 
In the future, we plan to investigate factors that influence LLM screening performance between secondary studies, explore avenues for per-study adaptations to improve individual secondary study screening accuracy with LLMs, and examine the potential of combining multiple LLMs with voting mechanisms to enhance the results. We conclude that guidelines are needed for the content and structure of secondary study research artifacts.

\section*{Acknowledgment}

The first and third author have been supported by the Strategic Research Council of Research Council of Finland (Grant ID 358471) and the second author has been funded by the Killam Postdoctoral Fellowship. We thank the OpenAI Research Access Program for providing us access to their API.

\section*{Data availability}

The research artifact for the study is available in Zenodo~\cite{zenodo_2_anon}.

\balance
\bibliographystyle{ieeetr}
\bibliography{ieee_manuscript}

\end{document}